\documentstyle[12pt]{article}
\titlepage

\title{Locally Metric Spacetimes}
\author{  Richard Atkins \\
        richard.atkins@twu.ca \\
		Department of Mathematics\\ 
		Trinity Western University \\
		7600 Glover Road \\
		Langley, BC, V2Y 1Y1 Canada}
\date{}
\newtheorem{fact}{Fact}

\newtheorem{lemma}[fact]{Lemma}
\newtheorem{proposition}[fact]{Proposition}
\newtheorem{theorem}[fact]{Theorem}
\newtheorem{corollary}[fact]{Corollary}
\begin{document}
\maketitle
\begin{abstract}
Spacetimes have conventionally been described by a global Lorentzian
metric on a differentiable four-manifold. Herein we explore
the possibility of spacetimes defined by a connection, which is locally but
not globally Levi-Civita. The general method of obtaining such 
connections is presented for the non-degenerate case followed by an
example that modifies the Robertson-Walker spacetimes for flat
spacelike hypersurfaces.
\end{abstract}

\newpage

\section{Introduction}
Since its inception, general relativity has comprised two principal ingredients:  
a four-dimensional manifold $M$ and a Lorentzian metric 
defined on $M$. Other objects, such as gauge or matter fields may also be employed, 
serving to influence the geometry of gravitation through the field equations
\begin{eqnarray} \label{fieldeqns}
 R_{\mu \nu} -\frac{1}{2} Rg_{\mu \nu} = \frac{8\pi G}{c^{4}} T_{\mu \nu} 
\end{eqnarray}
The four-dimensional manifold hypothesis has been the more flexible of the
two key constituents of spacetime; Kaluza and Klein, for instance,
have contemplated spaces of higher dimension,
whose compactifications were thought to explain some of the features of the physical world
(cf. \cite{hh}, \cite{ii}). 
On the other hand, the notion of a globally defined Lorentzian metric has remained, 
by and large, unmodified. 

The field equations  (\ref{fieldeqns}) are covariant with respect to the diffeomorphism
group {\it Diff}, of $M$. Early on, diffeomorphism covariance posed difficulties 
in the 
interpretation of general relativity, leading to the "hole" argument and its 
resolution by means of defining the gravitational state as a diffeomorphism class of 
solutions to (\ref{fieldeqns}). For certain systems, such as the massless scalar field
or the vacuum, the energy-momentum tensor is invariant under the group {\it Conf},  of
constant conformal rescalings of the metric: 
$g \longmapsto cg$, $c\in \Re^{+}$ (cf. \cite{ff}). 
Equations (\ref{fieldeqns}) are then also invariant
under such transformations and so physical states would be defined as equivalence
classes with respect to the enlarged group {\it Diff} $\times$ {\it Conf} (cf. \cite{aa}).

These considerations suggest the possibility of spacetimes that are described by a 
locally metric connection. A connection $\nabla$ on $M$ is 
{\it locally metric} if for each point $m \in M$ there exists
an open neighbourhood $U$ of $m$ and a metric on $U$ that is parallel with 
respect to $\nabla$; these locally defined parallel metrics might not piece together
into a global metric on $M$ (cf. \cite{bb}). 
We shall be primarily concerned with such connections that are, moreover, {\it non-degenerate}
in the sense that for each open subset $U$ of $M$ the vector space of parallel metrics on
$U$ has dimension at most one; otherwise certain pathologies may arise
with regard to the causal structure of spacetime. 
When the energy-momentum tensor is invariant under constant rescalings of the metric, 
the field equations (\ref{fieldeqns}) retain meaning in this context. 
In this case, the prospect arises that spacetime might be defined locally and not through
a global Lorentzian structure.

In the following section we develop a
characterization of non-degenerate, locally metric connections.
It is shown that such connections $\nabla$ satisfy an equation 
$\nabla h = h\otimes \Psi$, where $h$ is a metric on $M$ and $\Psi$ is a closed 1-form. 
Furthermore, $\nabla$ is globally metric if and only if $\Psi$ is exact.
The final section  investigates two examples. The first applies the methods 
of Section 2 for constructing non-globally metric spacetimes and the second examines
a degenerate locally metric solution.

\section{Locally Metric Connections} 
In this section we seek to determine the general structure of symmetric, non-degenerate, 
locally metric connections on a manifold $M$. They are 
described most succinctly in terms of a pair $(h,\Psi)$ where $h$ is a metric  and
$\Psi$ is a closed 1-form on $M$. From a practical perspective it is sometimes convenient to 
construct such connections by geometric considerations on the universal cover of $M$. 
We shall pursue both avenues below and the relation between them.

Let $\nabla$ be a non-degenerate, locally metric Lorentzian connection on a manifold $M$ of
dimension greater than two. At this stage we do not assume that $\nabla$ is symmetric.
By non-degeneracy, the set of local parallel metrics for $\nabla$ generates a line bundle 
$\pi:E\longrightarrow M$ over $M$. 
We may obtain a global non-vanishing section $h$ of $E$ by means of a partition of unity
argument. Let ${\cal U}=\{ U_{a}: a\in A \}$ 
be an open cover of $M$ such that $E_{|_{U_{a}}}$ is a trivial line
bundle over $U_{a}$ for each $a \in A$. Then on each $U_{a}$ there is a non-vanishing 
section $h_{a}: U_{a} \longrightarrow E_{|_{U_{a}}}$ with signature $(-1,+1,...,+1)$. 
Consider any two such local sections $h_{a_{1}}$ and $h_{a_{2}}$  that are defined at some
common point $m\in U_{a_{1}}\cap U_{a_{2}} \subseteq M$. 
Since $E$ has rank one, $h_{a_{1}}(m)$ and $h_{a_{2}}(m)$ are non-zero 
multiples of each other. Furthermore, since  $dim$ $M \geq 3$ and 
$h_{a_{1}}(m)$ and $h_{a_{2}}(m)$ have the same signature, they must be {\it positive}
multiples of each other. 

Let
$\{ \ell_{b}: b\in B \}$ with $\sigma:B \longrightarrow A$ be a partition of unity 
subordinate to ${\cal U}$: $supp( \ell_{b}) \subseteq U_{\sigma(b)}$ for each $b\in B$ and
$\sum_{b\in B} \ell_{b} = 1$. Consider the sum 
\[ h := \sum_{b\in B} \ell_{b}h_{\sigma(b)}\]
Given any  $m\in M$ there are only finitely many indices $b_{1},...,b_{n} \in B$
for which $\ell_{b_{i}}(m) \neq 0$. As we have observed, the bilinear forms 
$h_{\sigma(b_{1})}(m),...,h_{\sigma(b_{n})}(m)$ are positive multiples of each other and
so 
\[ h(m) = \sum_{b\in B} \ell_{b}(m)h_{\sigma(b)}(m) = 
          \sum_{i=1}^{n} \ell_{b_{i}}(m)h_{\sigma(b_{i})}(m)\]
is a non-degenerate bilinear form of signature $(-1,+1,...,+1)$. 
It follows that $h$ is a non-vanishing section of $E$.

Since $\nabla$ is locally metric
there exists, for each point $m$ in $M$, an open neighbourhood $U_{m}$ of $m$ and a parallel
local section $g_{m}:U_{m} \longrightarrow E_{|_{U_{m}}}$ of signature $(-1,+1,...,+1)$. 
In view of the fact that $E$ has rank one, $g_{m}$ and the restriction $h_{m} := h_{|U_{m}}$ 
are conformally equivalent: 
$h_{m}=\mu_{m}g_{m}$ for some function $\mu_{m}:U_{m}\rightarrow \Re^{+}$. 
Taking the covariant derivative of $h_{m}$ gives
$\nabla h_{m} = h_{m} \otimes dlog\mu_{m}$ on $U_{m}$.  
Since $\{ U_{m}$: $m\in M \}$ is an open cover of $M$,  $\nabla h = h \otimes \Psi$ 
for some closed 1-form $\Psi$ on $M$; on $U_{m}$ $\Psi$ restricts to $dlog\mu_{m}$.

Even though we have focused on locally metric Lorentzian connections above it is clear that
the results apply to locally metric Riemannian connections as well. The theorem  
summarizes the discussion so far.

\begin{theorem} \label{theorem:main} \hspace{1in}  \\
Let $\nabla$ be a non-degenerate, locally metric connection, which is Lorentzian on a manifold 
$M$ of dimension greater than two or Riemannian on a manifold of any dimension. 
Then there exists a metric $h$ and a closed 1-form
$\Psi$ on $M$ such that $\nabla$ satisfies the equation
\[ \nabla h = h \otimes \Psi \]
\end{theorem} 

A partial converse to the theorem is given by the following proposition.

\begin{proposition} \label{prop:local}
Suppose $\nabla h = h \otimes \Psi$, where $h$ 
is a metric and $\Psi$ is a closed 1-form on $M$. Then $\nabla$ is locally metric.
If $\Psi_{|_{U}} = dlog\mu$ on an open subset $U$ of $M$, for some
$\mu :U \longrightarrow \Re^{+}$, then $\nabla \mu^{-1} h_{|_{U}} =0$.  
\end{proposition}
{\it Proof}: \\
Since $\Psi$ is a closed 1-form, for each $m\in M$ there exists an open neighbourhood
$U$ of $m$ such that $\Psi_{|_{U}} = dlog\mu$, for some $\mu:U\longrightarrow \Re^{+}$.
On $U$,
\[ \nabla\mu^{-1}h_{|_{U}} = -\mu^{-2}h_{|_{U}}\otimes d\mu + 
   \mu^{-1} h_{|_{U}}\otimes\Psi_{|_{U}} = 0 \]
$\Box$

\begin{proposition}
Suppose $\nabla h = h \otimes \Psi$, where $h$ 
is a metric and $\Psi$ is a closed 1-form on $M$. \\
(a) If $\Psi$ is exact then $\nabla$ is a (globally)
metric connection on $M$. \\ 
(b) If $\nabla$ is a non-degenerate (globally) metric connection on $M$ 
then $\Psi$ is exact.
\end{proposition}
{\it Proof}: \\
$(a)$
follows from Proposition \ref{prop:local} with $U=M$.  \\
$(b)$
Suppose that $\nabla$ is a non-degenerate (globally) metric connection:
$\nabla g = 0$ for some metric $g$ on $M$. 
$h$ is locally conformally equivalent to a local parallel metric, 
by Proposition \ref{prop:local}.
The non-degeneracy of $\nabla$ then implies that $h = \mu g$ for some 
$\mu:M \longrightarrow \Re-\{0\}$. Thus $\Psi = dlog |\mu|$ is an exact 1-form on $M$. 
\\ $\Box$

\begin{corollary}
A non-degenerate, locally metric connection satisfying $\nabla h = h\otimes \Psi$
is a (globally) metric connection if and only if $\Psi$ is exact.
\end{corollary}

\begin{proposition} Let $h$ be a metric and $\Psi$ a closed 1-form on $M$. Then
the pair $(h,\Psi)$ determines a unique symmetric connection $\nabla$ on $M$ by the equation
\[ \nabla h = h \otimes \Psi \]
\end{proposition}
{\it Proof}: \\
For each point $m$ in $M$ there exists a connected open neighbourhood $U_{m}$ and a
function $\mu_{m}:U_{m}\longrightarrow \Re^{+}$, determined up to a positive constant multiple,
such that $\Psi_{|_{U_{m}}} = dlog \mu_{m}$. Thus $(h,\Psi)$ defines a unique symmetric
connection $\nabla$ on $M$ by the equations $\nabla \mu_{m}^{-1}h = 0$, $m\in M$. 
The proposition follows from the observation that the equations $\nabla \mu_{m}^{-1}h = 0$,
$m\in M$, are equivalent to $\nabla h = h\otimes \Psi$.
\\ $\Box$

This construction generalizes the Levi-Civita connection; if $\Psi$ is exact, with
$\Psi = dlog\mu$ for some $\mu:M\longrightarrow \Re^{+}$, then the symmetric connection
defined by $(h,\Psi)$ is the Levi-Civita connection of $g=\mu^{-1}h$.

The symmetric connection $\nabla$ determined by a pair $(h, \Psi)$ 
is not necessarily non-degenerate. It is easy to create
degenerate examples by considering product manifold structures. However, $\nabla$ will
be non-degenerate for a {\it generic} choice of $(h, \Psi)$.

Suppose $\nabla$ is a symmetric, non-degenerate connection determined by both
$(h_{1},\Psi_{1})$ and $(h_{2},\Psi_{2})$. 
By Proposition \ref{prop:local}, $h_{1}$ and $h_{2}$ are locally conformally 
equivalent to local parallel metrics and so by non-degeneracy $h_{1}$ and $h_{2}$
are conformally equivalent to each other:
$h_{2}=\alpha h_{1}$ for some function  $\alpha:M \longrightarrow \Re-\{0\}$. 
The equations $\nabla h_{i} = h_{i} \otimes \Psi_{i}$,
for $i=1,2$, give $\Psi_{2} = \Psi_{1} + dlog|\alpha|$. On the other hand, $(h, \Psi)$ and
$(\alpha h, \Psi + dlog|\alpha|)$ define the same symmetric connection for any non-vanishing
function $\alpha$ on $M$.

For the purpose of constructing locally metric connections it is often 
illuminating to carry the
analysis to the universal cover $\rho:\widetilde{M}\longrightarrow M$ of $M$.
Let $(h,\Psi)$ determine the locally metric connection $\nabla$ on $M$.
Pull $\nabla$, $h$ and $\Psi$ back to $\widetilde{\nabla}:= \rho^{*}(\nabla)$,
$\tilde{h}:=\rho^{*}(h)$ and $\tilde{\Psi}:=\rho^{*}(\Psi)$ on 
$\widetilde{M}$. Then 
\[ \widetilde{\nabla}\tilde{h} = \tilde{h}\otimes \tilde{\Psi} \]
$\tilde{\Psi}$ is exact since it is a closed form on a simply-connected manifold and so there 
exists a function $\tilde{f}:\widetilde{M} \longrightarrow \Re$ such that 
$\tilde{\Psi} = d\tilde{f}$. 
Put $\tilde{g}:=exp(-\tilde{f})\tilde{h}$. $\widetilde{\nabla}$ is the Levi-Civita
connection of $\tilde{g}$: 
\[  \widetilde{\nabla}\tilde{g} = 0 \]

For any $m\in M$ there exists a connected
open neighbourhood $U_{m}$ of $m$ in $M$ such that
$\rho^{-1}(U_{m})$ is the disjoint union 
\[ \rho^{-1}(U_{m}) = \coprod_{x\in \rho^{-1}(m)}V_{x} \]
of sheets $x \in V_{x} \subseteq \widetilde{M}$ each naturally diffeomorphic to $U_{m}$. 
Given two points $x,y \in \rho^{-1}(m)$ there exists  
a diffeomorphism $\phi_{x,y}:V_{x} \longrightarrow V_{y}$ defined by
\[ \rho_{|_{V_{y}}} \circ \phi_{x,y} = \rho_{|_{V_{x}}} \]
In particular, $\phi_{x,y}(x) = y$. Now,
\[ \begin{array}{lll}
\phi^{*}_{x,y}(\tilde{\Psi}_{|_{V_{y}}}) & = & 
           \phi^{*}_{x,y}(\rho^{*}_{|_{V_{y}}}(\Psi))  \\
& = & (\rho_{|_{V_{y}}}\circ \phi_{x,y})^{*}(\Psi) \\
& = & \rho_{|_{V_{x}}}^{*} (\Psi) \\
& = & \tilde{\Psi}_{|_{V_{x}}} 
\end{array} \]
Since the open sets $V_{x}$ and $V_{y}$ are required to be connected, it follows that
\[ \phi^{*}_{x,y}(\tilde{f}_{|_{V_{y}}}) = \tilde{f}_{|_{V_{x}}} 
   + k_{x,y} \]
for some constant  $k_{x,y}\in \Re$.
Therefore
\[ \begin{array}{lll}
\phi^{*}_{x,y}(\tilde{g}\hspace{0in}_{|_{V_{y}}}) & = & 
   \phi^{*}_{x,y}(e^{-\tilde{f}}\tilde{h\hspace{0in}}_{|_{V_{y}}}) \\
& = &   \phi^{*}_{x,y}(e^{-\tilde{f}}\rho^{*}(h)_{|_{V_{y}}}) \\
& = &   \phi^{*}_{x,y}(e^{-\tilde{f}}\hspace{0in}_{|_{V_{y}}})
        (\rho_{|_{V_{y}}} \circ \phi_{x,y})^{*}(h) \\
& = &   e^{-\tilde{f}-k_{x,y}}\rho^{*}(h)_{|_{V_{x}}} \\
& = &   e^{-k_{x,y}}e^{-\tilde{f}}\tilde{h}\hspace{0in}_{|_{V_{x}}} \\
& = &   e^{-k_{x,y}}\tilde{g}\hspace{0in}_{|_{V_{x}}}
\end{array} \]

Define the manifold 
\[ {\cal M} := \{ (x,y) \in \widetilde{M}\times\widetilde{M} : \rho(x) = \rho(y) \} \]
We have shown that $\tilde{g}$ satisfies the following condition:
\begin{eqnarray} \label{property}
  \phi^{*}_{x,y}(\tilde{g})(x) = c(x,y)\hspace{0.02in} \tilde{g}(x) 
\end{eqnarray} 
for all $(x,y) \in {\cal M}$, where $c: {\cal M} \longrightarrow \Re^{+}$
is a {\it locally constant} function. 
As the notation suggests, this condition does not depend upon the choice of
connected open neighbourhoods $U_{m}$. 
We state this as a lemma.
\begin{lemma} \label{lemma1}
Let $\nabla$ be a locally metric connection on $M$ defined by a pair $(h,\Psi)$. 
Then $\nabla$ is the projection of the Levi-Civita connection $\widetilde{\nabla}$ of a 
metric $\tilde{g}$ on the universal cover $\widetilde{M}$ of $M$, which 
satisfies condition (\ref{property}).
\end{lemma}

Conversely, suppose that $\tilde{g}$ is a metric on $\widetilde{M}$ 
satisfying condition (\ref{property}). Clearly, its Levi-Civita connection $\widetilde{\nabla}$
projects down to a locally metric connection $\nabla$ on $M$. We shall demonstrate that 
the projected connection $\nabla$ is, in fact, 
defined by a pair $(h,\Psi)$. To this end we seek a function 
$\tilde{\mu}:\widetilde{M} \longrightarrow \Re^{+}$ for which 
\begin{eqnarray} \label{requirement}
\phi^{*}_{x,y}(\tilde{\mu}\tilde{g})(x) = \tilde{\mu}\tilde{g}(x) 
\end{eqnarray}
for all $(x,y) \in {\cal M}$.

Let $x,y,z \in \widetilde{M}$ be such that $\rho(x) = \rho(y) = \rho(z)$.
From (\ref{property}),
\[ \begin{array}{lll}
c(x,z) \tilde{g}(x) & = & \phi^{*}_{x,z}(\tilde{g})(x) \\
   & = & (\phi_{y,z}\circ \phi_{x,y})^{*}(\tilde{g})(x) \\
   & = & \phi_{x,y}^{*}((\phi_{y,z}^{*}(\tilde{g}))(x) \\
   & = & c(y,z)\phi_{x,y}^{*}(\tilde{g})(x) \\
   & = & c(x,y)c(y,z)\tilde{g}(x)
\end{array} \]
Therefore
\begin{eqnarray} \label{cycle}
c(x,y)c(y,z) = c(x,z)
\end{eqnarray}

For each $m\in M$ fix some sheet $V_{m}$ in $\rho^{-1}(U_{m})$. Define the function
$\tilde{\mu}_{m}: \rho^{-1}(U_{m}) \longrightarrow \Re^{+}$ by
\[ \tilde{\mu}_{m}(x) := c(x,z) \]
where $z$ is the unique element in $V_{m}$ for which $\rho(z) = \rho(x)$.
Suppose $\rho(x)=\rho(y)=\rho(z)\in U_{m}$, with $z\in V_{m}$. Then
\[ \begin{array}{lll}
\phi^{*}_{x,y}(\tilde{\mu}_{m}\tilde{g})(x) & = & 
  \tilde{\mu}_{m}(\phi_{x,y}(x))\phi^{*}_{x,y}(\tilde{g})(x) \\
& = &   \tilde{\mu}_{m}(y)c(x,y)\tilde{g}(x) 
       \hspace{0.5in} \mbox{by (\ref{property})}  \\
& = &   c(y,z)c(x,y)\tilde{g}(x) \\
& = &   c(x,z)\tilde{g}(x) \hspace{0.92in} \mbox{by (\ref{cycle})} \\
& = &   (\tilde{\mu}_{m}\tilde{g})(x)
\end{array} \]
That is,
\begin{eqnarray} \label{local}
\phi^{*}_{x,y}(\tilde{\mu}_{m}\tilde{g})(x)  =  (\tilde{\mu}_{m}\tilde{g})(x)
\end{eqnarray}

Therefore $\tilde{\mu}_{m}$ satisfies (\ref{requirement}), 
but is only defined on an open subset of 
$\widetilde{M}$. To find a positive, globally defined function $\tilde{\mu}$ we use a partition 
of unity $\{ \ell_{a}: a\in A \}$ subordinate to the cover $\{ U_{m}: m\in M \}$ of 
$M$: $supp\hspace{0.03in} (\ell_{a}) \subseteq U_{\sigma(a)}$, where
$\sigma : A \longrightarrow M$.
Put 
\[ \tilde{\mu}: = \sum_{a \in A} (\ell_{a}\circ\rho) \tilde{\mu}_{\sigma (a)} \]
Consider $x,y \in \widetilde{M}$ with $\rho(x) = \rho(y)$ and let $a_{1},...,a_{n}$ 
be the indices in $A$ for which $\ell_{a_{i}}(\rho(y)) \neq 0$. Then 
$\rho (x) = \rho (y) \in supp\hspace{0.03in} (\ell_{a_{i}}) \subseteq 
U_{\sigma (a_{i})}$, for $1\leq i \leq n$.
\begin{eqnarray*}
\phi^{*}_{x,y}(\tilde{\mu}\tilde{g})(x) & = &   
  \sum_{a \in A} \phi^{*}_{x,y}((\ell_{a}\circ\rho) \tilde{\mu}_{\sigma (a)}\tilde{g})(x) \\
& = &   
  \sum_{i=1}^{n} \phi^{*}_{x,y}
    ((\ell_{a_{i}}\circ\rho) \tilde{\mu}_{\sigma (a_{i})}\tilde{g})(x) \\
& = & 
  \sum_{i=1}^{n} (\ell_{a_{i}}\circ\rho)(y) 
    \phi^{*}_{x,y}(\tilde{\mu}_{\sigma (a_{i})}\tilde{g})(x) \\
& = &
 \sum_{i=1}^{n} (\ell_{a_{i}}\circ\rho)(y) 
    (\tilde{\mu}_{\sigma (a_{i})}\tilde{g})(x) 
    \hspace{0.5in} \mbox{by (\ref{local})} \\
& = & 
   \sum_{a \in A} ((\ell_{a}\circ\rho) \tilde{\mu}_{\sigma (a)}\tilde{g})(x) \\
& = & (\tilde{\mu}\tilde{g})(x) 
\end{eqnarray*}
as required. 

Therefore $\tilde{\mu}\tilde{g}$ projects to a metric $h$ on $M$:
\[  \tilde{\mu}\tilde{g} = \rho^{*}(h) \]
To be more precise, define $h$ by the pullback
\[ h(\rho(x)):= (\rho_{|_{V_{x}}})^{-1*}(\tilde{\mu}\tilde{g}(x)) \]
for any $x\in \widetilde{M}$. To show that this is well-defined,
consider $y\in \widetilde{M}$ such that $\rho(y) = \rho(x)$. Then 
\[ \begin{array}{lll}
(\rho_{|_{V_{x}}})^{-1*}(\tilde{\mu}\tilde{g}(x))  & = & 
  (\rho_{|_{V_{x}}})^{-1*}(\phi^{*}_{x,y}(\tilde{\mu}\tilde{g})(x)) \hspace{0.5in}
     \mbox{by (\ref{requirement})} \\
& = & 
  (\rho_{|_{V_{x}}})^{-1*}(\phi^{*}_{x,y}(\tilde{\mu}\tilde{g}(y))) \\
& = & 
  (\phi_{x,y} \circ (\rho_{|_{V_{x}}})^{-1})^{*}(\tilde{\mu}\tilde{g}(y)) \\
& = & 
   (\rho_{|_{V_{y}}})^{-1*}(\tilde{\mu}\tilde{g}(y))  
\end{array} \]
By similarly pulling back the  equation
\[ \widetilde{\nabla} (\tilde{\mu}\tilde{g}) = (\tilde{\mu}\tilde{g})
   \otimes dlog\tilde{\mu} \]
it follows that
\[ \nabla h = h \otimes \Psi \]
for some closed 1-form $\Psi$ on $M$. 
Applying the previous lemma we arrive at the theorem below.
\begin{theorem}
Let $\nabla$ be a symmetric connection on $M$. $\nabla$ is defined by a pair
$(h,\Psi)$ if and only if 
$\nabla$ is the projection of the Levi-Civita connection $\widetilde{\nabla}$
 of a metric $\tilde{g}$ on
the universal cover $\widetilde{M}$ of $M$, which satisfies  condition (\ref{property}).
\end{theorem}

\section{Examples}

In this section we consider two examples, both chosen for the purpose of illustration rather 
than physical realism. The first will apply the procedure for 
finding locally metric connections, developed in the previous section, to Robertson-Walker
spacetimes. The second highlights some of the unusual features that can arise in the
degenerate case.

\hspace{1in} \\
\hspace{-0.3in} {\bf Non-degenerate case} \hspace{0.05in} 
Consider $M= \Re \times S^{1} \times K$,
where $S^{1}$ is the 1-sphere or circle and $K$ is the Klein bottle. 
$\Re$ will parametrize time and $S^{1} \times K$ will describe the topology of spacelike
hypersurfaces. We begin by seeking the first de Rham cohomology group of $M$.
The simplest case of the Poincar\'{e} Lemma is $H_{de R}^{0}(\Re) = \Re$ and 
$H_{de R}^{1}(\Re) = 0$. 
Also, it is well known that $H_{de R}^{0}(S^{1}) = H_{de R}^{1}(S^{1}) = \Re$.
Recall that the Klein bottle is obtained from the rectangle by identifying the edges
in the manner indicated by the arrows in the following diagram:

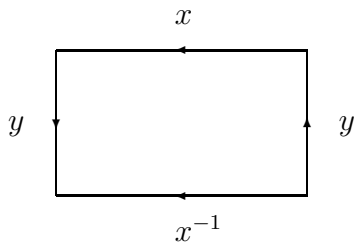
\begin{picture}(400,135)(-110,10)

\put(40,100){\line(1,0){45}}
\put(135,100){\vector(-1,0){50}}

\put(40,100){\vector(0,-1){30}}
\put(40,70){\line(0,-1){25}}

\put(40,45){\line(1,0){45}}
\put(135,45){\vector(-1,0){50}}

\put(135,100){\line(0,-1){25}}
\put(135,45){\vector(0,1){30}}

\put(85,110){$x$} 
\put(22,70){$y$}
\put(85,28){$x^{-1}$}
\put(147,70){$y$}

\end{picture} \\
The base point $k_{0}$ of $K$ corresponds to the corners of the rectangle, which
are all identified. 
By traversing the boundary of the rectangle in a counterclockwise direction we obtain a
loop based at $k_{0}$, which is homotopic, relative to the base point,
to the constant loop. Thus $xyx^{-1}y =1$ and the fundamental group of $K$ is 
\[ \pi_{1}(K) = Gp(x,y ; xyx^{-1}y) \]
that is, the group on two generators $x$ and $y$ subject to $xyx^{-1}y=1$.
The abelianization of the fundamental group is the abelian group generated by $x$ and $y$
restricted by the relation $x+y-x+y=0$, i.e. $2y=0$. Therefore
\[ \pi_{1}(K)/[\pi_{1}(K), \pi_{1}(K) ] \cong {\cal Z}\oplus {\cal Z}_{2}\]
The Hurewicz isomorphism then gives the first singular homology group of $K$ with integer 
coefficients: $H_{1}(K, {\cal Z}) \cong {\cal Z}\oplus {\cal Z}_{2}$. By a
Universal Coefficient theorem we obtain the first singular cohomology group with coefficients
in $\Re$:
\[ H^{1}(K,\Re)\cong Hom(H_{1}(K,{\cal Z}),\Re) \cong \Re\] 
According to de Rham's theorem, $H^{1}(K,\Re)$ is isomorphic to $H^{1}_{de R}(K)$. 
Hence 
\[ H^{1}_{de R}(K) \cong \Re \]
Also, since $K$ is connected, $H^{0}_{de R}(K) \cong \Re$. The K\"{u}nneth formula 
now provides the first de Rham cohomology of $M$:
\[ H_{de R}^{1}(M) \cong \bigoplus_{i+j+k =1}H_{de R}^{i}(\Re) \otimes 
                      H_{de R}^{j}(S^{1}) \otimes H_{de R}^{k}(K) \cong \Re^{2} \]  
    
$M$ may also be obtained from $\Re^{4}$ by the identifications
\[ (t,x,y,z) \sim (t, x+p, y,z) \sim (t, x, y+q ,-z) \sim (t,x,y, z+r) \]
where $p,q$ and $r$ are non-zero constants. 
$dx$ and $dy$ on $\Re^{4}$ are independent,
closed forms that project to $M$. Since $dim \hspace{0.03in} H_{de R}^{1}(M)=2$,
the general closed 1-form on $M$ is
\begin{eqnarray} \label{psi}
\Psi = -adx - bdy +dlog\alpha 
\end{eqnarray}
where $a$ and $b$ are constants and $\alpha$ is a positive function on $M$.  
The Robertson-Walker spacetime for flat spacelike hypersurfaces has the form
\[ ds^{2} = -dt^{2} + S^{2}(t)\Sigma \]
with $\Sigma = dx^{2}+dy^{2}+dz^{2}$. It projects to a Lorentzian metric $h$
on $M$. The pair $(h,\Psi)$ determines a unique symmetric connection $\nabla$ on
$M$ by the formula $\nabla h = h\otimes\Psi$. If $a=b=0$ then $\nabla$ is the Levi-Civita 
connection of a (global) metric on $M$.
 
From (\ref{psi}), $\Psi = dlog \mu$, for
\[ \mu = \alpha e^{-ax-by} \]
Therefore $\nabla$ is the Levi-Civita connection of the local metric
\begin{eqnarray} \label{localmetric}
 \alpha^{-1} e^{ax+by}(-dt^{2} + S^{2}(t)\Sigma) 
\end{eqnarray}
Let us consider the case $\alpha = 1$. On the spacelike hypersurfaces 
$t= t_{0}$, (\ref{localmetric}) reduces to $e^{ax+by}S^{2}(t_{0})\Sigma$,
which defines a locally Levi-Civita connection
$\nabla^{(3)}$. The non-zero Christoffel symbols for $\widetilde{\nabla}^{(3)}$,
the pullback of $\nabla^{(3)}$ to the universal cover $\Re^{3}$ of $S^{1}\times K$, are
\[ \begin{array}{llllll}
\Gamma^{i}_{xi} & = & \Gamma^{i}_{ix} & = & \frac{a}{2} & \hspace{0.3in} i=x,y,z \\
\Gamma^{i}_{yi} & = & \Gamma^{i}_{iy} & = & \frac{b}{2} & \hspace{0.3in} i=x,y,z \\
\Gamma^{x}_{ii} & = & -\frac{a}{2}    &   &             & \hspace{0.3in} i=y,z \\
\Gamma^{y}_{ii} & = & -\frac{b}{2}    &   &             & \hspace{0.3in} i=x,z \\
\end{array} \]
These are invariant with respect to the transformations
\[ \begin{array}{lll}
(x,y,z) & \longmapsto & (x+p,y,z) \\
(x,y,z) & \longmapsto & (x,y+q,-z) \\
(x,y,z) & \longmapsto & (x,y,z+r) 
\end{array} \]
verifying that $\nabla^{(3)}$ is, in fact, defined on $S^{1}\times K$. 

It is worth pointing out that a similar exploration performed on $M = \Re\times T^{3}$, 
where $T^{3}$ signifies the 3-torus, underscores some further characteristics analogous
to metric geometry. In this case, we find that given any two points on a spacelike 
hypersurface $t=t_{0}$, one can be mapped to the other by a translation on $T^{3}$, 
which preserves the locally metric connection. This naturally extends the notion of 
homogeneity to locally metric spacetimes.

\hspace{1in} \\
\hspace{-0.3in} {\bf Degenerate case} \hspace{0.05in} Consider the Lorentzian metric 
\[ ds^{2} = g_{1}+g_{2} \] 
on $\widetilde{M} :=\Re^{4}$, where
\[ g_{1}=  -cos\theta dt^{2} +2sin\theta dtd\theta  +cos\theta d\theta^{2} \]
and
\[ g_{2} = dx^{2} + dy^{2} \]  
$ds^{2}$ is the metric of a product of (pseudo-)Riemannian manifolds, so
the Levi-Civita connection $\widetilde{\nabla}$ of $ds^{2}$ is degenerate. 
Let $M$ be the quotient manifold of $\widetilde{M}$ with respect to the 
identifications
\[ (t,\theta,x,y) \sim (t,\theta+\pi,x,y) \]
Topologically, $M$ is the product of an infinite cylinder with the plane.
Since $g_{1}(t,\theta+\pi) = -g_{1}(t,\theta)$, the connection $\widetilde{\nabla}$ projects
down to a locally, but not globally, metric connection $\nabla$ on $M$; indeed,
the non-zero Christoffel symbols 
\[ \begin{array}{lllllllll}
\Gamma^{\theta}_{\theta\theta} & = & -\Gamma^{\theta}_{tt} & \hspace{-0.94in} = & 
       -\Gamma^{t}_{t\theta} & = &  -\Gamma^{t}_{\theta t}
               & \hspace{-0.3in} = & \frac{1}{2} sin\theta cos\theta \\
\Gamma^{t}_{tt} & = & -\Gamma^{\theta}_{t\theta} & \hspace{-0.94in} = & 
             -\Gamma^{\theta}_{\theta t} & = & -\frac{1}{2}sin^{2}\theta  \\
\Gamma^{t}_{\theta\theta} & = & -(cos^{2}\theta+\frac{1}{2}sin^{2}\theta) 
\end{array} \]
of $\widetilde{\nabla}$ are invariant with respect to the transformation 
$(t,\theta,x,y) \longmapsto (t,\theta +\pi ,x,y)$ on $M$.

$\nabla$ is not defined by $\nabla h = h\otimes \Psi$ for any pair $(h,\Psi)$.
Were it so, there would exist a positive function $\tilde{\mu}$ and a parallel metric 
$\tilde{g}$
on the universal cover $\widetilde{M}=\Re^{4}$ of $M$  such that $\tilde{\mu}\tilde{g}$ 
projects to $h$ on $M$. The most general metric on $\widetilde{M}$,
parallel with respect to $\widetilde{\nabla}$, is of the form
\begin{eqnarray} \label{metric}
\tilde{g} = ag_{1}+b_{1}dx^{2}+2b_{2}dxdy+b_{3}dy^{2}
\end{eqnarray}
for constants $a,b_{1},b_{2}$ and $b_{3}$ satisfying $a\neq 0$, $b_{1} >0$ and 
$b_{1}b_{3}-b_{2}^{2} > 0$.
If $\tilde{\mu}\tilde{g}$ were the pullback $\rho^{*}(h)$ of a metric $h$ on $M$
then we would have $\tilde{\mu}\tilde{g}(t,\theta+\pi,x,y) = 
\tilde{\mu}\tilde{g}(t,\theta,x,y)$, implying $\tilde{\mu}=0$.

Furthermore, the causal structure on $M$ is ambiguous. One might consider specifying a
causal structure about a point $m=(t,\theta ,x,y)$ on $M$ by a choice of parallel metric
in a neighbourhood of $m$, given by (\ref{metric}). 
Following this metric along
the loop $\gamma(s) = (t,\theta +s\pi,x,y)$, for $1\leq s\leq 1$,  
results in a second causal structure  about $m$ determined by 
\[ -ag_{1}+b_{1}dx^{2}+2b_{2}dxdy+b_{3}dy^{2} \]

\end{document}